\documentclass[twocolumn,floats,showpacs,amsmath,amssymb,prb]{revtex4}

\usepackage{graphicx}

\begin{document}

\title{Configurational entropy of hydrogen-disordered ice polymorphs}
\author{Carlos P. Herrero}
\author{Rafael Ram\'irez}
\affiliation{Instituto de Ciencia de Materiales de Madrid,
         Consejo Superior de Investigaciones Cient\'ificas (CSIC),
         Campus de Cantoblanco, 28049 Madrid, Spain }
\date{\today}

\begin{abstract}
The configurational entropy of several H-disordered ice polymorphs
is calculated by means of a thermodynamic integration along a path
between a totally H-disordered state and one fulfilling the
Bernal-Fowler ice rules.
A Monte Carlo procedure based on a simple energy model is used, so that
the employed thermodynamic path drives the system from high temperatures 
to the low-temperature limit.
This method turns out to be precise enough to give reliable values 
for the configurational entropy $s_{th}$ of different ice phases in the 
thermodynamic limit (number of molecules $N \to \infty$).
The precision of the method is checked for the ice model on a
two-dimensional square lattice.
Results for the configurational entropy are given for H-disordered
arrangements on several polymorphs, including 
ices Ih, Ic, II, III, IV, V, VI, and XII.
The highest and lowest entropy values correspond to ices VI and XII,
respectively, with a difference of 3.3\% between them.
The dependence of the entropy on the ice structures has been
rationalized by comparing it with structural 
parameters of the various polymorphs, such as the mean ring size.
A particularly good correlation has been found between the
configurational entropy and the connective constant derived from
self-avoiding walks on the ice networks.
\end{abstract}



\maketitle

\section{Introduction}

Water presents a large variety of solid phases. Up to now sixteen 
different crystalline ice polymorphs have been identified, most of them 
obtained by application of high pressures,
which produces a denser packing of water molecules than in
the usual hexagonal ice Ih.\cite{pe99,du10,ba12}
The determination of their crystal structures and
stability range in the pressure-temperature phase diagram has been
a subject of investigation over the last few decades.
However, in spite of the wide amount of experimental and theoretical 
works on the solid phases of water, some of their properties still lack 
a full understanding. 
This is largely due to the presence of hydrogen bonds between adjacent 
molecules, which give rise to some peculiar characteristics of these
condensed phases (the so-called `water anomalies').\cite{ei69,pe99,ro96}

In the known ice polymorphs (leaving out ice X), water molecules 
appear as well defined entities forming a network linked by H-bonds. 
In such a network each H$_{2}$O molecule is H-bonded to four others 
in a disposition compatible with the so-called Bernal-Fowler ice
rules. These rules state that each molecule is oriented so that
its two H atoms point toward adjacent oxygen atoms 
and that there must be exactly one H atom between two contiguous
oxygens.\cite{be33} We will refer to these rules simply
as `ice rules.'

Orientational disorder of the water molecules is present in
several ice phases. Oxygen atoms show full occupancy ($f$) of their 
crystallographic sites ($f = 1$), but hydrogen atoms may present a 
disordered distribution, as indicated by a fractional occupancy of their 
lattice positions ($f < 1$).
For example, hexagonal ice Ih presents full hydrogen disorder compatible 
with the ice rules (occupancy of H-sites $f = 0.5$).
Other polymorphs such as ice II are H-ordered, while some phases as
ices III and V display partial hydrogen order
(some fractional occupancies $f \neq 0.5$).

The sixteen known crystalline ice phases can be classified according to 
their network topology. Thus, one finds polymorphs sharing the same topology, 
as happens for ices Ih and XI. 
Something similar occurs for other ice polymorphs, and at present
six such pairs of ice structures are known:
Ih-XI, III-IX, V-XIII, VI-XV, VII-VIII, and XII-XIV,\cite{sa11} 
where the first polymorph in each pair is H-disordered and
the second is H-ordered.
In each case, both polymorphs share the same network topology and
are related one to the other through an order/disorder phase
transition.
In addition to these twelve polymorphs, there exist other ice structures 
for which no pair has been found:
ice Ic (H-disordered), ice II (H-ordered), and ice IV (H-disordered).
Finally, ice X is topologically equivalent to ices VII and VIII, with 
the main difference that in
ice X hydrogen atoms are situated on the middle point between adjacent 
oxygen atoms, so that water molecules lose in fact their own entity. 
Then, one has nine topologically different ice structures.
We note that the ice VII network (as well as ices VIII and X)
consists of two independent interpenetrating subnetworks not
hydrogen-bonded to each other (each of them equivalent to the ice 
Ic network).
There is another case, the pair VI-XV, for which the network is also
made up of two disjoint subnetworks, but they are not equivalent to
the network of any other known ice polymorph.

The first evaluation of the configurational entropy $S$ associated to
H disorder in ice was presented by Pauling,\cite{pa35}
who found $S = N k_B \ln (3/2)$ for a crystal of $N$ water molecules.
This result turned out to be in rather good agreement with 
the `residual' entropy obtained from the experimental heat capacity 
of ice Ih,\cite{gi36,ha74} even though the calculation did not
consider the actual structure of this ice polymorph.
Nagle\cite{na66} calculated later the residual entropy of hexagonal 
ice Ih and cubic ice Ic by a series method, and found for both
polymorphs very similar values, which were close to but slightly higher 
than the Pauling's estimate. 
In recent years, Berg {\em et al.}\cite{be07,be12} have employed 
multicanonical simulations to calculate the configurational entropy of 
ice Ih, assuming a disordered H distribution consistent with the ice rules.
Furthermore, the configurational entropy of partially ordered ice
phases has been studied earlier,\cite{ho87,ma04b,be07b} and will not be 
addressed here.

Besides their application to condensed phases of water, ice-type models 
are relevant for other types of solids displaying atomic or spin 
disorder,\cite{is04,po13} as well as in statistical-mechanics
studies of lattice problems.\cite{zi79,be89}
At present, no analytical solutions for the ice model in actual
three-dimensional (3D) ice structures are known, so that
no exact values for the configurational entropy are available.
An exact solution was found by Lieb\cite{li67,li67b} for 
the two-dimensional (2D) square lattice, where the entropy resulted
to be $S = N k_B \ln W$, with $W = (4/3)^{3/2} = 1.5396$, a value 
somewhat higher than Pauling's result $W_P = 1.5$.

A goal of the present work consists in finding relations
between thermodynamic properties and topological characteristics of ice
polymorphs.
In this line, it has recently been shown that the network topology
plays a role to quantitatively describe the configurational entropy
associated to hydrogen disorder in ice structures.\cite{he13}
This has been known after the work by Nagle,\cite{na66} who showed
that the actual structure is relevant for this purpose, by comparing
his results for ices Ih and Ic with Pauling's value for the configurational 
entropy, which is in fact the limiting value corresponding to a loop-free 
network (Bethe lattice).

In this paper, we present a simple but `formally exact' numerical 
method, to obtain the configurational entropy of H-disordered ice 
structures.  It is based on a thermodynamic integration from high to low
temperatures for a model whose lowest-energy states are consistent 
with the ice rules.
The configurational space is sampled by means of Monte Carlo (MC) 
simulations, which provide us with accurate values for the heat capacity 
of the considered model.
This method has been employed earlier to calculate the entropy of
H-disordered ices Ih and VI in Ref.~\onlinecite{he13}.
Apart from ice Ih, Ic, and VI, we are not aware of any calculation of 
the configurational entropy of other H-disordered ice polymorphs. 
Entropy calculations have been also carried out earlier for H distributions
displaying partial ordering on the available crystallographic sites,
as happens for ices III and V.\cite{ma04b}

\section{Method of calculation}

For concreteness, we recall the ice rules, as established by
Bernal and Fowler\cite{be33} in 1933:
First, there is one hydrogen atom between each pair of neighboring 
oxygen atoms, forming a hydrogen bond, and second,
each oxygen atom is surrounded by four H atoms, two of them 
covalently bonded and two other H-bonded to it.  
Fulfillment of these rules implies that short-range order must be present 
in the distribution of H atoms in ice polymorphs.

Calculating the configurational entropy of different ice
structures, assuming that the hydrogen distribution is only conditioned
by the ice rules, is equivalent to find the number of possible
hydrogen configurations compatible with these rules.  
Since the entropy is a state function, one can employ any thermodynamic
integration that connects the required state with any other state whose 
thermodynamic properties are known.  For the present problem, 
this can be adequately done by considering a simple
model, in which an energy function is defined so that all states
compatible with the ice rules are equally probable (have the same
energy $U_0$), and other hydrogen configurations have higher energy.
The calculation of the configurational entropy of ice is thus reduced to 
finding the degeneracy of the ground state of considered model.
Taking the H configurations according to their Boltzmann
weight, as is usual in the canonical ensemble, only those with the
minimum energy $U_0$ will appear for $T \to 0$.
In this respect, the only requirement for a valid energy model is that 
it has to reproduce the ice rules at low temperature. Thus, irrespective 
of its simplicity, it will yield the actual configurational 
entropy if an adequate thermodynamic integration is carried out. 

Our model is defined as follows. We consider an ice structure as defined 
by the positions of the oxygen atoms, so that each O atom has four
nearest O atoms. This defines a network, where the nodes are the O
sites, and the links are H-bonds between nearest neighbors. 
The network coordination is four, which gives a total of $2N$ links,
$N$ being the number of nodes. We assume that on each link there is
one H atom, which can move between two positions 
on the link (close to one oxygen or close to the other).
We will use reduced variables, so that quantities such as the energy $U$ 
and the temperature $T$ are dimensionless. 
The entropy per site $s$ presented below is related with the 
physical configurational entropy $S$ as $S = N k_B s$, $k_B$ being
Boltzmann's constant.

\begin{figure*}
\vspace{0.3cm}
\hspace{-0.2cm}
\includegraphics[width=12.0cm]{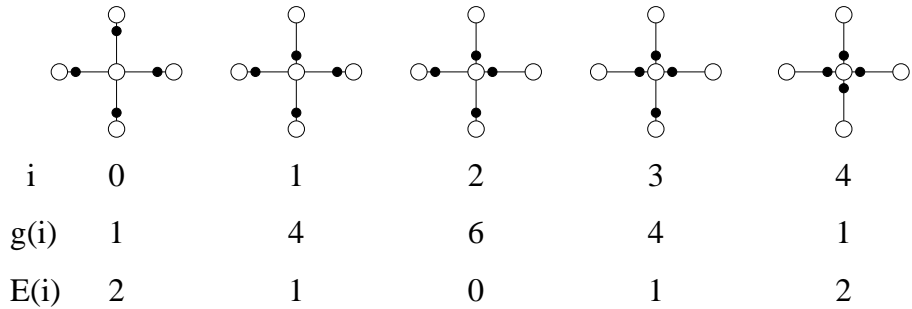}
\caption{
Schematic illustration of the different hydrogen configurations around
an oxygen atom. Open and solid circles represent oxygen and
hydrogen atoms, respectively.
For each configuration, $i$ is the number of H atoms close
(covalently bonded) to the central oxygen,
$g(i)$ indicates its multiplicity, and
$E(i)$ is its dimensionless energy in the model described in the text.
}
\label{f1}
\end{figure*}

For a given configuration of H atoms on an ice network, the energy $U$ 
is defined as:
\begin{equation}
U = \sum_{n=1}^N  E(i_n)
\end{equation}
where the sum runs over the $N$ nodes in the network,
and $i_n$ is the number of hydrogen atoms covalently bonded to the
oxygen on site $n$, which can take the values $i_n$ = 0, 1, 2, 3, or 4.
The energy associated to site $n$ is defined as $E(i_n) = |i_n - 2|$ 
(see Fig.~1), having a minimum for $i_n = 2$ ($E_2 = 0$), which 
imposes accomplishment of the ice rules at low temperature.
In this way, all hydrogen configurations compatible with the ice
rules on a given structure are equally probable, because they
have the same energy ($U_0 = 0$ in our case).

 This energy model is a convenient tool to derive the configurational 
entropy, but it does not represent any realistic interatomic
interaction. In fact, we are not dealing with an actual ordering process
in an H-bond network, but with a numerical approach to `count' 
H-disordered configurations. 
One can obtain the number of configurations compatible with 
the ice rules on a given structure by a thermodynamic integration, going
from a reference state ($T \to \infty$) for which the H configuration is 
random (does not follow the ice rules), to a state in which these rules 
are strictly satisfied ($T \to 0$). 

This model assumes that all hydrogen arrangements that satisfy the
Bernal-Fowler rules are energetically degenerate. In the actual ice 
structures there will be variations in the energy due to different 
hydrogen bonding patterns.\cite{kn06,si12}
For large system size and at temperatures at which the H-disordered
phases are stable, the crystal structures
corresponding to different hydrogen arrangements will form a nearly
degenerate energy band, populated almost uniformly.
Thus, the idea underlying the calculation of `residual' entropies is that
the accessible hydrogen configurations do not correspond to the
actual stable phase at $T = 0$ (which should be ordered), but to an
equilibrium H-disordered phase at higher $T$.
This means that small energy differences between different hydrogen
arrangements are supposed to not affect significantly their relative weight
for the residual entropy, as these arrangements will have nearly the same
probability in the (equilibrium) disordered phase.

We obtain the configurational entropy per site at temperature $T$ from the 
heat capacity, $c_v$, as
\begin{equation}
 s(T) = s(\infty) + \int_{\infty}^T  \frac{c_v(T')}{T'} \, d \, T'
\label{st1}
\end{equation}
where
\begin{equation}
  c_v(T) = \frac{1}{N} \frac{d \langle U \rangle}{d \, T}  \, ,
\label{cvt}
\end{equation}
and the entropy in the high-temperature limit ($T \to \infty$) is 
given by
\begin{equation}
  s(\infty) = \frac{1}{N} \ln (2^{2 N}) = 2 \ln 2
\label{entinfi}
\end{equation}
Here, $2 N$ is the number of links in the considered network, and
$2^{2 N}$ is the total number of possible configurations in our model, 
as two different positions are accessible for an H atom on each link.

Although Eq.~(\ref{st1}) is exact in principle, a practical problem
in this thermodynamic integration appears because the limit 
$T \to \infty$ cannot be reached in actual simulations.
One can introduce a high-temperature cutoff $T_c$ for this 
integration, but even if $T_c$ is very high (in our units $T_c \gg 1$), 
it may introduce uncontrolled errors in the calculated entropies.
Thus, a reasonable solution to this problem can be calculating the
integral in Eq.~(\ref{st1}) up to a temperature $T_c$, and then
obtaining the remaining part (from $T_c$ to $T = \infty$) by an
alternative method. 
With this purpose, an analytical procedure has been introduced in
Ref.~\onlinecite{he13}, and is described in the following section
dealing with an independent-site model.

\subsection{Independent-site model}

For a real ice network, thermodynamic variables at high temperatures 
($T \gg 1$ here) can be well described by considering `independent' nodes, 
as in Pauling's original calculation for a hypothetical loop-free network.
In such a case, the partition function can be written as
\begin{equation}
  Z =  \frac {z^N} {2^{2N}}  \,  ,
\label{zz}
\end{equation} 
$z$ being the one-site partition function, and the term $2^{2N}$ in the
denominator appears to avoid counting links with zero or two H atoms.
This correction is in fact similar to that introduced by Pauling in his 
entropy calculation.\cite{pa35}

To obtain the partition function $z$, we consider the 16 possible
configurations of four hydrogen atoms, as given in Fig.~1, which yields 
at temperature $T$:
\begin{equation}
  z =   \sum_{i=0}^4 g(i) \,  \exp \left[ - \frac{E(i)}{T} \right]   
    =   6 + 8 \, {\rm e}^{-1/T} + 2 \, {\rm e}^{-2/T}  \, .
\label{z68}
\end{equation}
From this expression for $z$ one obtains the partition function $Z$
using Eq.~(\ref{zz}).  Thus, the high-temperature limit of $Z$ is 
$Z_{\infty} = 16^N / 4^N = 2^{2N}$, which
is the number of configurations compatible with the condition of
having an H atom per link, as indicated above.

In this independent-site model, the energy per site at temperature $T$ is
\begin{equation}
 \langle u \rangle =  \frac1N \, T^2 \, \frac{\partial \ln Z}{\partial T} =
       \frac{4}{z} \left( 2 \, {\rm e}^{-1/T} + {\rm e}^{-2/T} \right) \, .
\label{umean}
\end{equation}
For the entropy per site we have
\begin{equation}
   s  =  \frac1N \, \frac {\partial \, (T \ln Z)} {\partial \, T} \, ,
\label{s1}
\end{equation}
and using Eqs.~(\ref{zz}) and (\ref{umean}), one finds
\begin{equation}
   s  =  \ln z + \frac {\langle u \rangle}{T} - 2 \ln 2   \, .
\label{s2}
\end{equation}
In the limit $T \to 0$ (ice rules strictly fulfilled),
we have
\begin{equation}
   \lim_{T \to 0} \frac {\langle u \rangle}{T} = 0  \, ,
\label{lim0}
\end{equation}
and $z \to 6$ [see Eq.(\ref{z68})], so that
$s(0) = \ln \frac32$, which coincides with Pauling's 
result.\cite{pa35}

For our calculations on real ice networks we will employ a cutoff
temperature  $T_c = 10$. Introducing this temperature in Eq.~(\ref{s2}), 
we find for the independent-site model $s(T_c)$ = 1.38414, which 
coincides with the value obtained in Ref.~\onlinecite{he13} from a series 
expansion in powers of $1/T$ up to fourth order.

\subsection{Actual ice networks}

 The above independent-site equations are exact for a loop-free network,
i.e., the so-called Bethe lattice or Cayley tree.\cite{zi79,be89}
For real ice structures, the corresponding networks
contain loops, which means that the factorization of the partition
function in Eq.~(\ref{zz}) is not possible.  Nevertheless,
at high temperatures thermodynamic variables for the real structures
converge to those of the independent-site model, and this is the fact
we use to obtain the high-temperature part of our thermodynamic 
integration.

Thus, for the actual ice networks we calculate the configurational 
entropy per site for hydrogen-disordered distributions compatible with
the ice rules (accomplished in our method for $T \to 0$) as:
\begin{equation}
 s(0) = s(T_c) - \int_0^{T_c}  \frac{c_v(T')}{T'} \, dT'
\label{st2}
\end{equation}
where $s(T_c)$ = 1.38414 is the value obtained for the independent-site
model, and the heat capacity $c_v$ is derived from MC simulations
for the considered ice polymorph in the temperature range from $T = 0$
to $T = T_c$ (we will take $T_c = 10$ here).


\begin{table}[ht]
\vspace{-0.2cm}
\caption{Crystal system and space group for the ice polymorphs
studied in this work, along with the references from where
crystallographic data were taken. $N_{max}$ is the number of water
molecules in the largest supercell employed here for each ice
structure.\\}
\centering
\setlength{\tabcolsep}{7pt}
\vspace*{0.2cm}
\begin{tabular}{ccccc}
   Phase & Crystal system & Space group &  Ref. &  $N_{max}$  \\[2mm]
   \hline 
 Ih    & Hexagonal    & $P6_3$/$mmc$, 194 &  \onlinecite{pe57} & 2880 \\
 Ic    &   Cubic      & $Fd\bar{3}m$, 227 &  \onlinecite{ko44} & 2744 \\
 II    & Rhombohedral & $R\bar{3}$, 148   &  \onlinecite{ka71} & 2592 \\
 III   & Tetragonal   & $P4_12_12$, 92    &  \onlinecite{lo00} & 2592 \\
 IV    & Rhombohedral & $R\bar{3}c$, 167  &  \onlinecite{en81} & 2400 \\
 V     & Monoclinic   & $A2/a$, 15        &  \onlinecite{lo00} & 2688 \\
 VI    & Tetragonal   & $P4_2/nmc$, 137   &  \onlinecite{ku84} & 3430 \\
 XII   & Tetragonal   & $I\bar{4}2d$, 122 &  \onlinecite{lo98} & 2400 \\
   &&& \vspace*{-0.3cm} \\
\end{tabular}
\label{tb:ice_crystal_systems}
\end{table}

Using the energy model described above, we carried out
MC simulations on ice networks of different sizes.
Crystallographic data employed to generate the ice structures are
given in Table~I, along with the corresponding references from where
they were taken.
In this Table we also present the size $N_{max}$ of the largest
supercells employed for the different ice polymorphs.
Details of the simulations are similar to those employed earlier
for ices Ih and VI.\cite{he13}
For the hydrogen distribution on the available sites along the MC
simulations, we assumed periodic boundary conditions.
Sampling of the configuration space at different temperatures was
carried out by the Metropolis update algorithm.\cite{bi10}
For each network we considered 360 temperatures in the range
between $T$ = 10 and $T$ = 1, and 200 temperatures in the interval from
$T$ = 1 to $T$ = 0.01.
For each considered temperature, we carried out $10^4$ MC steps
for system equilibration, followed by $8 \times 10^6$ steps for
averaging of thermodynamic variables. A MC step included
an update of $2 N$ (the number of H-bonds) hydrogen positions
successively and randomly chosen.
Once calculated the heat capacity $c_v(T)$ at the considered
temperatures, the integral in Eq.~(\ref{st2}) was numerically
evaluated by using Simpson's rule.
Finite-size scaling was employed to obtain the configurational
entropy per site $s_{th}$ corresponding to the thermodynamic limit
(extrapolation to infinite size, $N \to \infty$).

Although the heat capacity $c_v(T)$ can be obtained from MC
simulations by numerical differentiation using directly its definition
in Eq.~(\ref{cvt}), we found more practical to derive it from the energy
fluctuations at temperature $T$.
Thus, we employed the expression\cite{ch87}
\begin{equation}
   c_v(T) = \frac {(\Delta U)^2} {N \, T^2}  \,  ,
\end{equation}
where $(\Delta U)^2 = \langle U^2 \rangle - \langle U \rangle^2$.
In particular, using this expression we obtained $c_v$ values with
smaller statistical noise than employing Eq.~(\ref{cvt}).

Now a relevant question is whether the high-temperature replacement of 
real ice structures by the independent-site model introduces any 
appreciable error in the calculated configurational entropies.
We have checked this point in two different ways. First, we  have
compared for some temperatures higher than $T_c$ the energy and heat
capacity obtained in the independent-site model with those derived from
MC simulations for actual ice structures.
Second, we have calculated the configurational entropy obtained with our
procedure in the low-$T$ limit for the ice model on a 2D square lattice, 
for which an exact analytical solution is known.\cite{li67}
Both procedures indicate that the error of our method is smaller than
the error bars associated to the statistical uncertainty of the MC
simulations and the error due to the numerical integration of the
ratio $c_v/T$ in Eq.~(\ref{st2}). This is discussed below.

\begin{figure}
\vspace{-1.1cm}
\hspace{-0.5cm}
\includegraphics[width= 9cm]{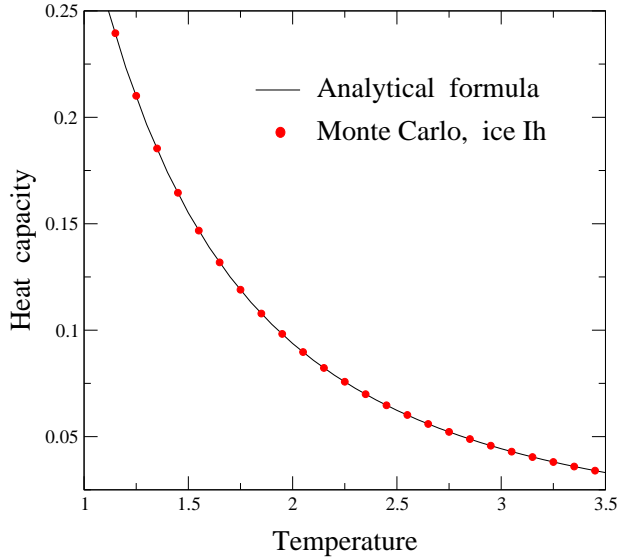}
\vspace{-0.8cm}
\caption{
Heat capacity per site as a function of temperature.
Solid circles are data points obtained from MC simulations for
ice Ih.
Error bars are less than the symbol size.
The solid line represents the analytical function obtained for the
independent-site model as $c_v(T) = d \langle u \rangle / d T$ with
$\langle u \rangle$  given by Eq.~(\ref{umean}).
}
\label{f2}
\end{figure}

In Fig.~2 we present the heat capacity $c_v(T)$ in the temperature
region from $T$ = 1 to 3.5. Symbols are results of our MC simulations
for ice Ih, whereas the solid line is the analytical result obtained for
the independent-site model as $c_v(T) = d \langle u \rangle / d T$, with
$\langle u \rangle$  given in Eq.~(\ref{umean}).
Data obtained by both procedures coincide within the statistical error
bars of the simulation results. Although not appreciable in the figure,
we observe a trend of the $c_v$ values derived from MC simulations to be
slightly larger than the analytical ones at temperatures $T \sim 1$.
At temperatures $T \sim T_c$, differences between both methods are
clearly less than the error bars.

\begin{figure}
\vspace{-1.1cm}
\hspace{-0.5cm}
\includegraphics[width= 9cm]{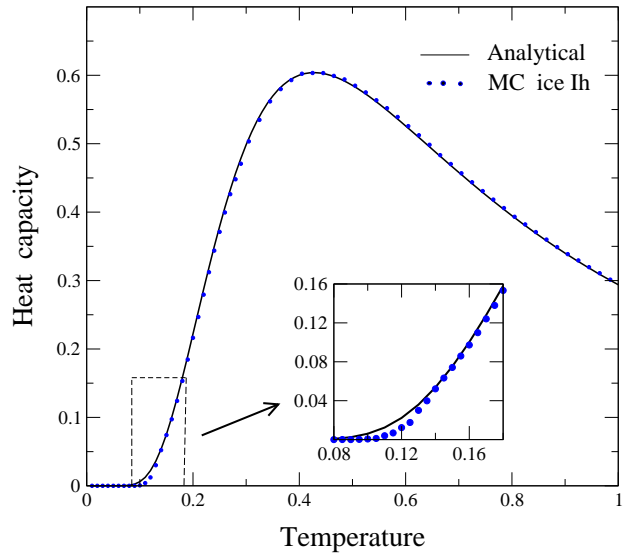}
\vspace{-0.8cm}
\caption{
Heat capacity per site as a function of temperature.
Solid circles are data points obtained from simulations for the
structure of ice Ih.
Error bars are less than the symbol size.
The solid line corresponds to the independent-site model:
$c_v(T) = d \langle u \rangle / d T$ with
$\langle u \rangle$  given by Eq.~(\ref{umean}).
}
\label{f3}
\end{figure}

The heat capacity reaches a maximum at $T_{max} \sim 0.4$, and with our
numerical precision vanishes at temperatures $T < 0.08$.
Larger differences between both procedures are observed at temperatures
$T < 1$, as shown in Fig.~3. At $T \gtrsim 0.4$, results of the MC
simulations are slightly larger than the analytical ones, and the
opposite happens at $T < 0.4$.
We observe in the inset that an appreciable difference appears in the 
region between $T$ = 0.10 and 0.14.
Note that the area under the curve $c_v(T)$ is independent of the 
considered network, since it is given by the difference
between the high- and low-temperature limits of the energy:
\begin{equation}
  \int_0^{\infty} c_v(T) \, d T =
     \lim_{T \to \infty} \langle u \rangle  -  \frac{U_0}{N} =  0.75
\end{equation}
For the energy model considered here we have $U_0 = 0$.

An alternative procedure to calculate the configurational entropy can 
consist in directly obtaining the density of states as a function of the 
energy, as described elsewhere for order/disorder problems in 
condensed-matter problems.\cite{he92}
Thus, the entropy can be obtained in the present problem from the
number of states with $U = 0$, i.e., compatible with the ice rules.

To avoid any possible misunderstanding, we note that the simple model
used in our calculations is not aimed at reproducing any physical
characteristic of ice polymorphs (as order/disorder transitions) beyond 
calculating the entropy of H distributions compatible with the
Bernal-Fowler rules.
This model implicitly takes for granted that the distribution of H atoms 
on an ice network does not necessarily display long-range order, and only 
requires strict fulfillment of the ice rules (short-range order) for
$T \to 0$.  The configurational entropy derived in this way is what has 
been traditionally called residual entropy of ice.\cite{pa35}
Taking all this into account, we are not allowing for any violation of the 
third law of Thermodynamics, which implies that for $T \to 0$ an ice
polymorph displaying H disorder cannot be the thermodynamically stable
phase. Thus, it is generally accepted that the equilibrium ice
polymorphs in the low-temperature limit show ordered proton structures, 
as expected from a vanishing of the entropy. In this context,
order-disorder transitions have been observed between several pairs of
ice phases.\cite{du10,ba12,si12} These transitions are accompanied by
an orientational reorganization of water molecules, which turns out to be 
a kinetically unfavorable rearrangement of the H-bond network.

\section{Results}

We have applied the integration method described above to calculate the
configurational entropy of the ice model in eight structures.
These structures include the ice polymorphs showing hydrogen
disorder, but also the network of ice II, for which no hydrogen disorder
has been observed. In this case, the results presented below have to be
understood to apply to a hypothetical structure with the same topology
as ice II.
Moreover, for ices III and V, a partial hydrogen disorder has been found
(site occupancies $f$ different from 0.5), a question that will not be
taken into account here (see the discussion below on this question).
Ice VII will not be discussed, since it consists of two sub-networks, each
of them equivalent to the ice Ic network (see the Introduction).

For each considered network, we obtained the configurational
entropy $s_N$ in the limit $T \to 0$ for several supercells of size 
$N$, as described in Sect.~II.
Then, the entropy per site in the thermodynamic limit, $s_{th}$,
is obtained by extrapolating $N \to \infty$.
 The precision of our procedure can be checked by
calculating the configurational entropy for the ice model on a
2D square lattice, for which an exact analytical solution 
is known.\cite{li67,li67b}    

\begin{figure}
\vspace{-1.1cm}
\hspace{-0.5cm}
\includegraphics[width= 9cm]{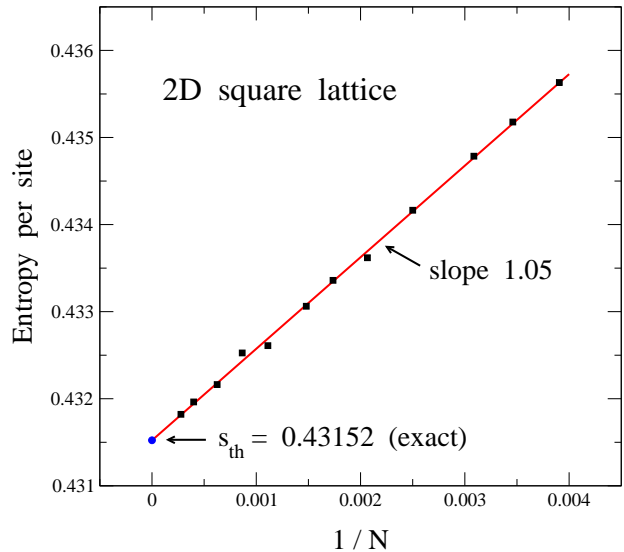}
\vspace{-0.8cm}
\caption{
Entropy per site $s_N$ vs inverse lattice size ($1/N$) for the
2D square lattice.
Solid squares indicate results of our thermodynamic integration in the
limit $T \to 0$.  Error bars are in the order of the symbol size.
A solid circle shows the exact analytical result obtained by
Lieb in the limit $N \to \infty$.\cite{li67}
}
\label{f4}
\end{figure}

In Fig.~4 we present $s_N$ for the 2D square lattice 
as a function of the inverse supercell size $1/N$.  Solid squares 
represent the entropy values yielded by our thermodynamic integration. 
It is found that the entropy per site $s_N$ decreases for increasing 
size $N$, and a linear dependence of $s_N$ on $1/N$ is observed for 
$N \gtrsim 150$, as
\begin{equation}
  s_N = s_{th} + \frac{\alpha}{N} \, ,
\label{sns}
\end{equation}
with a slope $\alpha = 1.05$. 
For supercell sizes $N < 150$ (not displayed in Fig.~4), $s_N$ values
slightly deviate from the linear trend given in Eq.~(\ref{sns}), 
being smaller than those predicted from the fit for $N > 150$.
Extrapolating $s_N$ to infinite size ($1/N \to 0$) gives a value
$s_{th}$ = 0.43153(3), in agreement with the exact solution for the 
square lattice derived by Lieb\cite{li67} from a transfer-matrix
method: $s_{th} = \frac32 \ln (4/3) = 0.43152$.

\begin{figure}
\vspace{-1.1cm}
\hspace{-0.5cm}
\includegraphics[width= 9cm]{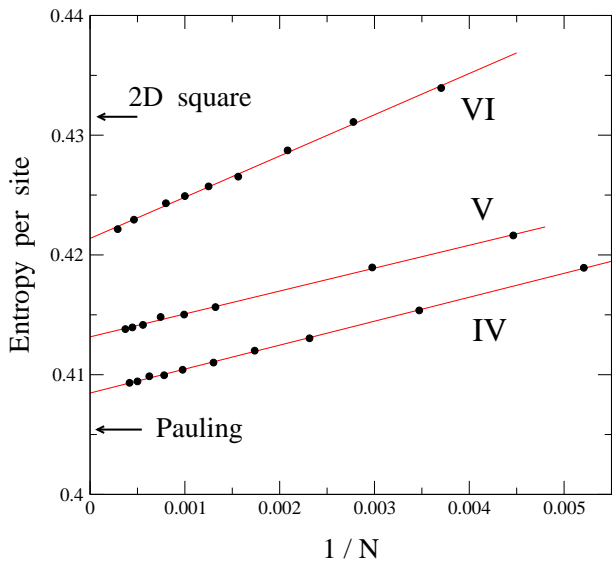}
\vspace{-0.8cm}
\caption{
Entropy per site $s_N$ as a function of the inverse supercell size
($1/N$) for three ice structures: IV, V, and VI.
Data points were derived from thermodynamic integration in the limit
$T \to 0$.  Error bars are less than the symbol size.
Arrows indicate the entropy $s_{th}$ corresponding to the 2D square
lattice and the Pauling result.
}
\label{f5}
\end{figure}

For the 3D structures of ice polymorphs we also found a
linear dependence of the configurational entropy $s_N$ on the 
inverse supercell size.
This is shown in Fig.~5 for ices IV, V, and VI. 
One observes that both the slope $\alpha$ and the infinite-size limit
$s_{th}$ depend on the ice structure. 
The slope is clearly larger than in the 2D case ($\alpha = 1.05$),
as for ices IV, V, and VI we find $\alpha$ = 2.00(2), 1.91(3), and 
3.44(6), respectively.  For comparison, arrows in Fig.~5 indicate 
the entropy per site for the 2D square lattice
and the Pauling result. The data for the ice polymorphs converge in the
infinite-size limit to $s_{th}$ between those two values. 


\begin{table*}
\caption{Configurational entropy $s_{th}$ associated to hydrogen disorder
in several ice polymorphs, as derived from our Monte Carlo simulations.
For comparison we also give results for the 2D square
lattice, as well as Pauling's value (Bethe lattice).
$W = \exp(s_{th})$;
$n_p$: number of data points employed in the linear fits;
$\alpha$: slope of the linear fit as in Eq.~(\ref{sns});
$\rho$: correlation coefficient.
}
\label{tab:2}
\vspace{3mm}

\begin{tabular}{c c c c c c}
   Ice  &  $n_p$  & $\alpha$  &  $s_{th}$ &  $W$  &  $\rho$   \\[2mm]
\hline  \\[-2mm]
  Ih   &  9   &  1.84(2)  &  0.41069(8)  &  1.50786(12)  &  0.9994
\\[2mm]
  Ic   &  10  &  1.82(2)  &  0.41064(7)  &  1.50778(11)  &  0.9994
\\[2mm]
   II     & 12   &  1.87(1)  &  0.40919(4)  &  1.50560(6)   &  0.9998
\\[2mm]
   III    & 12   &  1.79(2)  &  0.41165(7)  &  1.50931(11)  &  0.9990
\\[2mm]
   IV     & 10   &  2.00(2)  &  0.40847(4)  &  1.50451(6)   &  0.9997
\\[2mm]
   V      &  8   &  1.91(3)  &  0.41316(6)  &  1.51159(9)   &  0.9992
\\[2mm]
   VI     &  9   &  3.44(6)  &  0.42138(11) &  1.52406(16)  &  0.9990
\\[2mm]
  XII     & 10   &  2.08(3)  &  0.40813(5)  &  1.50400(8)   &  0.9992
\\[2mm]
\hline  \\[-2mm]
  Square  & 12   &  1.05(1)  &  0.43153(3)  &  1.53961(5)   &  0.9994
\\[2mm]
 Pauling  &  -   &      -      &  0.405465    &  1.5 &  - \\
  \hspace{1.5cm} &   \hspace{1.5cm} &  \hspace{2.5cm} & \hspace{2.5cm} &
   \hspace{2.5cm} & \hspace{2.5cm}  \\
\end{tabular}
\end{table*}

For other ice structures not shown in Fig.~5 we also obtained $s_{th}$ 
values between the square lattice and the Pauling result (see Table~II).
The largest value corresponds to ice VI ($s_{th} = 0.42138(11)$), and
the smallest one was found for ice XII ($s_{th} = 0.40813(5)$).
Between both values one finds a relative difference of about 3\%.
For hexagonal ice Ih, the configurational entropy is found to be an 1.3\%
larger than Pauling's estimate.
For other ice polymorphs, the difference between $s_{th}$ and the Pauling
result ranges from 0.7\% for ice XII to 3.9\% for ice VI. 

Information on the least-square fit of $s_N$ values for different ice
polymorphs, such as the parameter $\alpha$, the correlation coefficient
$\rho$, and the number of points $n_p$ employed in the fit, are given in
Table~II. 
When discussing the configurational entropy of ice,
it is usually given in the literature the parameter $W$, instead of
the entropy $s_{th}$ itself.
Here, we find directly $s_{th}$ from thermodynamic integration and
finite-size scaling, so that $W$ values given in Table~II were calculated 
for the different ice polymorphs as $W = \exp (s_{th})$.
The error bars $\Delta W$ and $\Delta s_{th}$ for the values of
$s_{th}$ and $W$ are related by the expression
$\Delta W = W \, \Delta s_{th}$, as can be derived by differentiating
the exponential function.
Error bars for $s_{th}$ represent one standard deviation, as
given by the least-square procedure employed to fit the data.
These error bars are less than $\pm 10^{-4}$, except for ice VI, for
which we found $\Delta s_{th} = 1.1 \times 10^{-4}$. 

For comparison with the results found for the ice structures, we also 
give in Table~II data for the 2D square lattice (linear fit shown in
Fig.~4) and for a loop-free network with coordination $z = 4$
(Pauling's result).
Note that, although the result obtained by Pauling was intended
to reproduce the residual entropy of ice Ih, it does not take into
account the actual ice network, but only the fourfold coordination of 
the structure. 

Now the question arises why the entropy per site decreases for 
increasing supercell size. A qualitative explanation is the following.
Let us call $\Omega_N$ the number of configurations compatible with 
the ice rules for supercell size $N$. For two
independent cells of size $N$, the number of possible configurations
is then $\Omega_N^2$. If one puts both cells together to form a
larger cell of size $2 N$, one has $\Omega_{2N} < \Omega_N^2$,
since configurations not fitting correctly at the border between both 
$N$-size cells are unsuitable and have to be rejected.
Then, one has 
\begin{equation}
  s_{2N} = \frac{1}{2N} \, \ln \Omega_{2N} <
          \frac{1}{2N} \, \ln \Omega_{N}^2 = 
          \frac{1}{N} \, \ln \Omega_{N} = s_N
\end{equation}
Something similar can be argued in general for two different cell sizes, 
$N < M$, so that $s_M < s_N$.
This means that the decrease in $s_N$ for rising $N$ can be
considered to be  a consequence of the cyclic boundary conditions.
Assuming that $s_N$ behaves regularly as a function of $1/N$ in the
thermodynamic limit ($1/N \to 0$), one can write for large $N$ the  
power expansion: $s_N = s_{th} + \alpha/N + \beta/N^2 + ...$.
In fact, this is what we found from our simulations for the different
ice polymorphs, with the parameter $\beta$ so small that a 
linear dependence of $s_N$ on $1/N$ is consistent with
the results of the thermodynamic integration at least for $N > 150$.

For ice Ih we find $s_{th}$ = 0.41069(8).
As already observed in the analytical result by Nagle\cite{na66} and
in the multicanonical simulations by Berg {\em et al.},\cite{be07,be12}
the configurational entropy is higher than the earlier estimate by
Pauling.
Our result is somewhat higher than that of Nagle,\cite{na66}
who found $s_{th}$ = 0.41002(10),
and slightly higher than the results of multicanonical simulations,
although the latter may be compatible with our data, taking into
account the error bars. We do not find at present a clear reason why
our result for ice Ih is higher than that found by Nagle using a series
method. The error bar given by this author is not statistical, and one 
could argue that maybe it was underestimated, but this question should
be further investigated.
A discussion on the various results for ice
Ih is given in Ref.~\onlinecite{he13}.

For cubic ice Ic, we found $s_{th}$ = 0.41064(7), which coincides
within error bars with our result for hexagonal ice Ih.
To our knowledge, the only earlier calculation for ice Ic is that
presented by Nagle in 1966,\cite{na66} who concluded that within the
limits of his estimated error, the entropy is the same for ices Ic
and Ih, i.e., $s_{th}$ = 0.41002(10).
We reach the same conclusion concerning both ice polymorphs, but in
our case the entropy value is higher by an 0.15\%.
This difference seems to be small, but it is significant as it
amounts to more than eight times our error bar (standard deviation).

It is known that ices III and V show partially ordered hydrogen
distributions.
This means that some fractional occupancies of H-sites are
different from 0.5. 
Then, the actual configurational entropy for these polymorphs is
lower than that corresponding to a H-disordered distribution compatible 
with the ice rules.
MacDowell {\em et al.}\cite{ma04b} calculated the configurational
entropy of partially ordered ice phases by using an analytical
procedure, based on a combinatorial calculation.
For ices III and V, these authors found entropy values $s_{th}$ = 0.3686 
and 0.3817, respectively, which they used to calculate the phase diagram
of water with the TIP4P model potential.
Taking the Pauling result as the entropy value for H-disordered
ice phases, the data by MacDowell {\em et al.} mean an entropy
reduction of 9.1\% and 5.9\%, for ice III and V, respectively.
This is an appreciable entropy reduction, which is noticeable in the
calculated phase diagram.\cite{ma04b}
However, if one takes for the H-disordered phases more realistic
entropy values, as those derived here, the entropy drop due to partial
hydrogen ordering is somewhat larger. In fact, one finds 10.5\% for
ice III and 7.6\% for ice V.
In any case, it is known that the hydrogen occupancies of the
different crystallographic sites may change to some extent with 
the temperature, so that the configurational entropy will also
change.\cite{ma04b}

\section{Relation with structural properties}

 Topological characteristics of solids at the atomic or molecular
level have been considered along the years to study properties of 
various types of materials.\cite{pe72,li85,we86}
For crystalline ice polymorphs, a discussion of different network 
topologies and the relation of ring sizes in the various phases with 
the crystal volume was presented by Salzmann {\em et al.}\cite{sa11}, 
and more recently by the present authors in Ref.~\onlinecite{he13b}.
As indicated earlier,\cite{he13} one expects that structural and
topological aspects of ice networks can be relevant to understand
quantitatively the configurational entropy of H-disordered structures.
In particular, this entropy should in some way be related with
effective parameters describing topological aspects of the
corresponding networks.


\begin{table}[ht]
\vspace{-0.2cm}
\caption{Minimum ($L_{min}$), maximum ($L_{max}$), and mean ring size
$\langle L \rangle$ for different ice structures, along with the
corresponding parameter $a$ and connective constant $\mu$.
$a$ is the coefficient of the quadratic term in the coordination
sequence, $M_k \sim a k^2$, as in Eq.~(\ref{mk}).
}
\vspace{0.3cm}
\centering

\setlength{\tabcolsep}{10pt}
\begin{tabular}{c c c c c c}
 Network & $L_{min}$  &  $L_{max}$  &  $\langle L \rangle$  &
         $a$   &   $\mu$      \\[2mm]
\hline  \\[-2mm]
 Ih         &  6  &   6  &   6    &  2.62  &   2.8793   \\[2mm]

 Ic         &  6  &   6  &   6    &  2.50  &   2.8792   \\[2mm]

 II         &  6  &  10  &  8.52  &  3.50  &   2.9049   \\[2mm]

 III        &  5  &   8  &  6.67  &  3.24  &   2.8714   \\[2mm]

 IV         &  6  &  10  &  9.04  &  4.12  &   2.9162   \\[2mm]

 V          &  4  &  12  &  8.36  &  3.86  &   2.8596   \\[2mm]

 VI         &  4  &   8  &  6.57  &  2.00  &   2.7706   \\[2mm]

 XII        &  7  &   8  &  7.6   &  4.26  &   2.9179   \\[2mm]
\hline  \\[-2mm]

 Square     &  4  &   4  &   4    &  0.0   &   2.6382   \\[2mm]

 Bethe      &  -- &  --  &  --    &   --   &   3.0000   \\[2mm]
\end {tabular}
\label{tb:coord_seq}
\end{table}

 The tendency of the entropy $s_{th}$ to increase due to the presence 
of loops in the ice structures has been pointed out recently.\cite{he13}
 The Pauling approximation ignores the presence of loops, as
happens in the Bethe lattice,\cite{zi79,be89} and yields a value 
$s_{th} = 0.40547$.
For ice Ih, which contains six-membered rings of water molecules,
the configurational entropy is higher than Pauling's value by 
an 1.3\%, and it is still higher for the ice model on a 2D square 
lattice, including four-membered rings (a 6.4\% with respect to the 
Pauling approach).
For other ice polymorphs with larger ring sizes, one can expect
a smaller configurational entropy for disordered hydrogen
distributions, and thus closer to the Pauling result.
In Table~III we give the minimum ($L_{min}$), maximum ($L_{max}$), and
mean ring size $\langle L \rangle$ for the ice structures considered
here.

\begin{figure}
\vspace{-1.1cm}
\hspace{-0.5cm}
\includegraphics[width= 9cm]{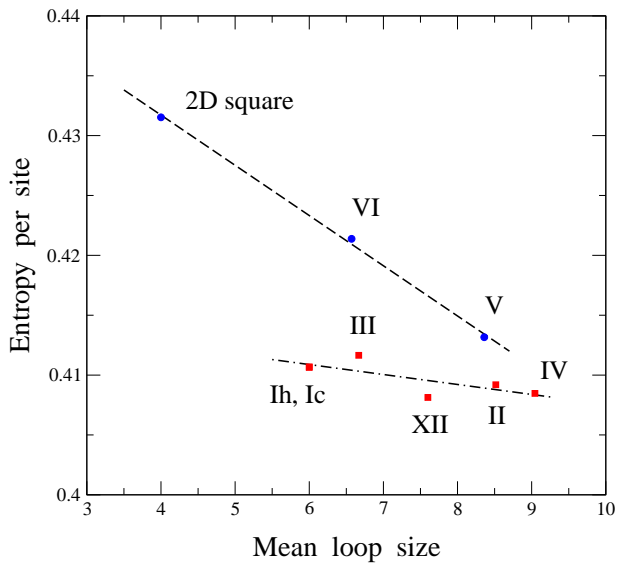}
\vspace{-0.8cm}
\caption{
Configurational entropy per site $s_{th}$ vs mean loop size
$\langle L \rangle$ for several ice structures.
Error bars for the entropy are less than the symbol size.
The 2D square lattice is included for comparison.
The dashed line is a least-square linear fit to the data of
structures containing four-membered rings ($L_{min} = 4$).
The dashed-dotted line is a fit for the ice structures with
$L_{min} > 5$.
}
\label{f6}
\end{figure}

In this line, we consider the mean loop size $\langle L \rangle$ as 
a first topological candidate to be compared with $s_{th}$.
A correlation between both quantities was suggested in
Ref.~\onlinecite{he13}, and will be considered in more detail here.
In Fig.~6 we present the configurational entropy of various ice 
polymorphs vs $\langle L \rangle$.
For comparison, we also include a data point for the 2D square lattice, 
where all loops have $L = 4$.
There appears a tendency of the configurational entropy
$s_{th}$ to decrease for rising $\langle L \rangle$, but one observes
a large dispersion of the data points for different ice polymorphs, 
indicating a low correlation between both quantities.
However, a careful observation of the data displayed in Fig.~6 reveals
that points corresponding to the ice structures are roughly 
aligned, with the exception of ices V and VI, which are located away from 
the general trend. A common characteristic of these two ice polymorphs 
is that they contain four-membered rings ($L_{min} = 4$).
In fact, we find a good linear correlation between $s_{th}$ and
the mean ring size $\langle L \rangle$ for these structures and the
2D square lattice (dashed line in Fig.~6).
The dashed-dotted line is a linear fit to the data points 
of ice structures not including four-membered rings. For these structures
we have $L_{min} > 4$, i.e. $L_{min} = 5$ for ice III,
$L_{min} = 6$ for ices Ih, Ic, II, and IV, and
$L_{min} = 7$ for ice XII.
We observe that the point corresponding to ice III appears above the 
dashed-dotted line, whereas that corresponding to ice XII lies below
it. This appears to be in line with the trend found for ices V and VI, 
since $L_{min}({\rm III}) < 6 < L_{min}({\rm XII})$, indicating that
for a given value of $\langle L \rangle$, a lower $L_{min}$ causes a
larger entropy $s_{th}$.

These observations allow us to
conclude that the configurational entropy tends to lower as the mean ring 
size increases, but the presence of four-membered rings greatly affects 
the actual value of the entropy $s_{th}$. Thus, the configurational entropy 
for ice VI turns out to be clearly higher than that of ice III, even though
both polymorphs have similar values of $\langle L \rangle$.
However, the effect of four-membered rings is smaller for larger
$\langle L \rangle$, as can be seen by comparing $s_{th}$ values for
ices II an V (see Fig.~6).

Further than structural rings,
the topology of a given ice network can be characterized by the so-called
coordination sequences $\{M_k\}$ ($k$ = 1, 2, ...), where $M_k$
is the number of sites at a topological distance $k$ from a reference
site (we call topological distance between two sites the number of bonds 
in the shortest path connecting one site to the other).\cite{he13b}
For 3D structures, $M_k$ increases at large distances as:
\begin{equation}
         M_k \sim a \, k^2     \hspace{3mm}  ,
\label{mk}
\end{equation}
where $a$ is a network-dependent parameter. $M_k$ increases
quadratically with $k$ just as the surface of a sphere increases
quadratically with its radius.
For structures including topologically non-equivalent sites, the actual
coordination sequences corresponding to different sites may be
different, but the coefficient $a$ coincides for all sites in a given
structure.\cite{he13b}
This parameter $a$ can be used to define a `topological density' $\rho_t$ 
for ice polymorphs as $\rho_t = w \, a$, where $w$ is the number of
disconnected subnetworks in the considered network.
Usually $w = 1$, but for ice structures including two interpenetrating
networks (as ices VI and VII) one has $w = 2$.
The parameter $a$ and the topological density $\rho_t$ of crystalline ice
structures have been calculated earlier.\cite{he13b}
It was found that $a$ ranges from 2.00 (ice VI) to 4.27 (ice XII), as
can be seen in Table~III for the ice polymorphs studied here.

\begin{figure}
\vspace{-1.1cm}
\hspace{-0.5cm}
\includegraphics[width= 9cm]{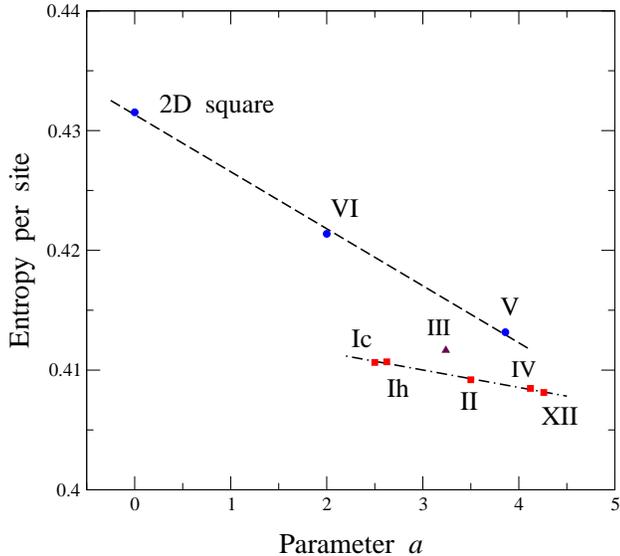}
\vspace{-0.8cm}
\caption{
Configurational entropy per site $s_{th}$ vs parameter $a$ of
coordination sequences for several ice structures.
Error bars for the entropy are less than the symbol size.
A point for the 2D square lattice ($a = 0$) is shown  for comparison.
The dashed line is a least-square linear fit to the data of
structures containing four-membered rings ($L_{min} = 4$).
The dashed-dotted line is a fit for the ice structures
with $L_{min} > 5$.
}
\label{f7}
\end{figure}

In Fig.~7 we display the configurational entropy per site $s_{th}$ of 
ice polymorphs vs the coefficient $a$ derived from the corresponding
coordination sequences.
We include also a point for the 2D square lattice, for
which $a = 0$, as in this case the dependence of $M_k$ on the
distance $k$ is linear: $M_k = 4 k$.
We find that $s_{th}$ decreases for increasing parameter $a$, but
there appears an appreciable  dispersion in the data points, which 
can be associated to the size of structural  rings in the considered ice 
polymorphs.       In particular,
 ice networks including four-membered rings ($L_{min} = 4$)
depart clearly from the general trend for the other structures,
similarly to the fact observed above in Fig.~6 for the relation
between entropy and mean loop size.
Thus, two different fits are also displayed in Fig.~7. 
The dashed line is a linear fit for the ice polymorphs with
$L_{min} = 4$ (ices V and VI), along with the 2D square lattice. 
The dashed-dotted line is a least-square linear fit for the structures 
with $L_{min} > 5$, with all of them having $L_{min} = 6$ except ice XII,
for which $L_{min} = 7$.
There remains ice III ($L_{min} = 5$), not included in either of
the fittings, whose data point (solid triangle) lies on the region
between both lines in Fig.~7.

From these results we conclude that the configurational entropy tends 
to decrease as the parameter $a$ (density of nodes) increases.
However, the presence of small rings in the ice structures strongly 
affects the actual value of $s_{th}$.  
Thus, $s_{th}$ for ices V and VI (with $L_{min} = 4$) is clearly larger 
than that corresponding to ice polymorphs with similar $a$ values but
with $L_{min} > 4$.

We see that the correlation of the configurational
entropy with both the mean loop size and the topological parameter $a$
is highly affected by the presence of structural rings with small
number of water molecules, in particular by those of the smallest size
found for the known crystalline polymorphs, $L = 4$.
It would be desirable to find other structural or topological
characteristics which could correlate more directly with the entropy
$s_{th}$. In this line, various parameters defined in statistical-mechanics
studies of lattice problems have been found earlier to be useful 
to characterize several thermodynamic properties.
One of these parameters, which can be used to characterize the topology
of ice structures, is the `connective constant' $\mu$, or effective
coordination number.\cite{do70,he95b,he14} This network-dependent 
parameter can be calculated from the long-distance behavior of the number 
of possible self-avoiding walks (SAWs) in the corresponding
structures.\cite{bi10,mc76,ra85}

 A self-avoiding walk on an ice network is defined as a walk in the 
simplified structure (only oxygen sites, see Sect.~II) which can never 
intersect itself.  The number $C_n$ of possible SAWs of length $n$ 
starting from a given network site is asymptotically 
given by\cite{pr91,mc76,ra85}
\begin{equation}
    C_n \sim   n^{\gamma - 1}   \mu^n  \hspace{2mm} ,
\label{cn}
\end{equation}
where $\gamma$ is a critical exponent which takes a value
$\approx 7/6$ for 3D structures.\cite{mc76,ca98,ch02} 
Then, the connective constant $\mu$ can be obtained as the limit
\begin{equation}
   \mu = \lim_{n \to \infty}  \frac{C_n}{C_{n-1}}   \, .
\label{mu}
\end{equation}
This parameter $\mu$ depends upon the particular topology of
each structure, and has been accurately determined for standard
3D lattices.\cite{mc76}
For the networks of different ice polymorphs, the connective constant
$\mu$ has been recently calculated, and was compared with other topological
characteristics of these networks.\cite{he14}
It ranges from 2.770 for ice VI to 2.918 for ice XII, as shown in 
Table~III.

\begin{figure}
\vspace{-1.1cm}
\hspace{-0.5cm}
\includegraphics[width= 9cm]{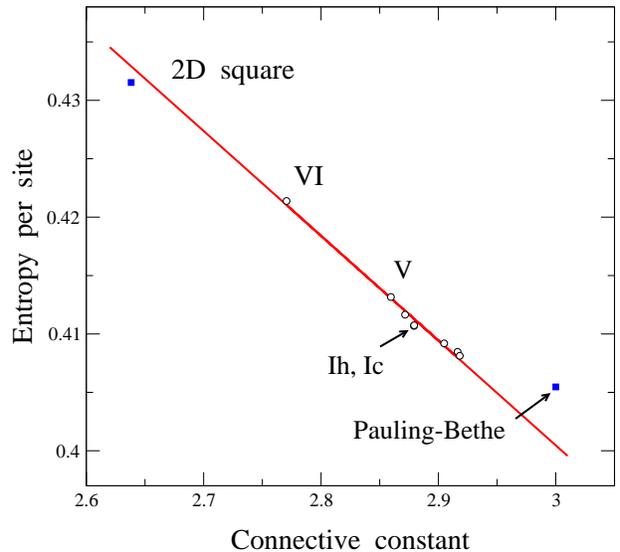}
\vspace{-0.8cm}
\caption{
Configurational entropy per site $s_{th}$ vs connective constant $\mu$
for several ice structures (open circles). Values for the 2D
square lattice and the Bethe lattice (Pauling model) are displayed for
comparison (solid squares). The solid line is a least-square linear fit
to the data of ice structures.
Error bars for the entropy are less than the symbol size.
}
\label{f8}
\end{figure}

In Fig.~8 we present the configurational entropy per site $s_{th}$ 
vs the connective constant $\mu$ for the considered ice polymorphs, 
as well as for the loop-free Bethe lattice (Pauling's result) and 
the 2D square lattice.
One observes in this figure a negative correlation between the entropy
$s_{th}$ and $\mu$. The line is a least-square fit to the data
points of the ice polymorphs.
Data corresponding to the square lattice and the Bethe lattice were
not included in the fit, as they depart from the trend of the ice
structures.
This seems to be due to the role played by dimensionality in the 
actual values of the configurational entropy
(the Bethe lattice has no well-defined dimension $d$, and for some
purposes it can be considered as $d \to \infty$).

In relation to the data displayed in Fig.~8 for ice polymorphs,
we cannot find at present any reason why the correlation between
$s_{th}$ and $\mu$ should be strictly linear, but it
is clear that there is a relation between both magnitudes.
To obtain some insight into this question, a qualitative argument is 
the following.
The connective constant $\mu$ is a measure of the mean number 
of sites connected to a node in long SAWs.
Then, for the distribution of H atoms on the links of 
the ice structures according to the Bernal-Fowler rules, a larger $\mu$ 
is associated to a rise in the correlations between occupancies of
different hydrogen sites, as information on the actual occupancy of a link
propagates `more effectively' through the ice network.\cite{he02b,he03c}
This correlation gives rise to a decrease
in the number of configurations compatible with the
ice rules for a given structure, which means a reduction of 
the configurational entropy of the hydrogen distribution on the
considered ice network.
Similar relations between SAWs and order/disorder problems, such as 
Ising or lattice-gas models, have been found and 
discussed earlier.\cite{pe02,do70,he95b}
In this line, the relation observed here between configurational
entropy and SAWs should be investigated further in the context of the
statistical mechanics of lattice models.

Going back to the topological characteristics of ice structures,
the smallest cycles of water molecules in crystalline ice phases are
four-membered rings. As observed in Figs.~6  and 7, and commented above,
the presence of these rings in ices V and VI imposes correlations
on the hydrogen atom distribution clearly stronger than in other ice phases
including larger rings.
A qualitative argument for this fact can be obtained from an estimation
of the way in which the presence of rings affects the configurational
entropy, similarly to calculations carried out by Hollins \cite{ho64} for
ice Ih. This author found that the correction to the Pauling value
($W = 1.5$) introduced by six-membered rings (considered to be
independent) is $\Delta W = 0.0041$. Using the same reasoning, with the
so-called `poles' in Ref.~\onlinecite{ho64}, we find for ice VI a correction
about 4.5 times larger ($\Delta W = 0.0185$).
These values of $\Delta W$ are smaller than those found in the more precise
thermodynamic integration given in Table.~II, as the former do
not take into account correlations such as those present between adjacent
rings and at larger distances in the network. In any case, it is clear
that the correlation introduced by
four-membered rings is significantly stronger than for larger rings.
This stronger correlation affects
similarly the configurational entropy and connective constant of the
H-bond networks studied here, as manifested by the correlation displayed
by both quantities (see Fig.~8), where all considered ice phases, containing
or not four-membered rings, follow the same trend.

\section{Concluding remarks}

We have presented results for the configurational entropy of
H-disordered ice structures, as calculated from a thermodynamic 
integration for a simple model which reproduces the Bernal-Fowler
rules in the limit $T \to 0$.  This model has allowed us to obtain
the entropy for different ices polymorphs, taking into account the
actual structure of each of them.
This procedure has turned out to be very precise, indicating that 
the associated error bars can be made relatively small without using
elaborate procedures, but only employing standard statistical mechanics 
and numerical techniques.
In fact, the estimated error bars of the calculated entropy values are 
about one part in 4,000, i.e. around a 0.03\% of the actual 
$s_{th}$ value. 

For real ice structures we found entropy values ranging from 
$s_{th}$ = 0.4081 for ice XII to $s_{th}$ = 0.4214 for ice VI, which 
means a difference of 3.3\% between these polymorphs.
These values are 0.7\% and  3.9\% larger than Pauling's result, 
respectively.
For ice Ih we obtained $s_{th}$ = 0.4107, which is close to the result 
of earlier calculations using multicanonical simulations.

Given the diversity of ice structures, the resulting values of the 
configurational entropy have been analyzed in view of different aspects
of the corresponding networks. 
In this line, it was found a trend of the entropy to lower as the mean 
ring size $\langle L \rangle$ increases. However, the presence of 
four-membered rings in some polymorphs (ices V and VI) markedly affects 
the actual value of the entropy. 
A similar conclusion can be derived for the correlation between
the configurational entropy and other topological parameters of ice
networks, such as the parameter $a$ giving the topological density
$\rho_t$.
A better correlation has been found between the configurational entropy 
of H-disordered ice polymorphs and the corresponding connective constants, 
derived from self-avoiding walks on their networks.
This allows one to rationalize the entropy differences obtained for 
the ice polymorphs in terms of parameters depending on the actual
topology of the associated structures.

The entropy difference found here for real ice polymorphs may be 
important for detailed calculations of the phase diagram of 
water,\cite{sa04,ra12b,ra13} as indicated earlier for ice phases 
with partial proton ordering.\cite{ma04b}
Moreover, the method described here to calculate the configurational
entropy of ice can be applied to other condensed-matter problems, 
such as the so-called spin-ice systems, where understanding the spin
disorder is crucial to explain their magnetic
properties.\cite{is04,po13}

\begin{acknowledgments}
This work was supported by Direcci\'on General de Investigaci\'on (Spain) 
through Grant FIS2012-31713, 
and by Comunidad Aut\'onoma de Madrid 
through Program MODELICO-CM/S2009ESP-1691.
\end{acknowledgments}


\begin{thebibliography}{54}
\expandafter\ifx\csname natexlab\endcsname\relax\def\natexlab#1{#1}\fi
\expandafter\ifx\csname bibnamefont\endcsname\relax
  \def\bibnamefont#1{#1}\fi
\expandafter\ifx\csname bibfnamefont\endcsname\relax
  \def\bibfnamefont#1{#1}\fi
\expandafter\ifx\csname citenamefont\endcsname\relax
  \def\citenamefont#1{#1}\fi
\expandafter\ifx\csname url\endcsname\relax
  \def\url#1{\texttt{#1}}\fi
\expandafter\ifx\csname urlprefix\endcsname\relax\def\urlprefix{URL }\fi
\providecommand{\bibinfo}[2]{#2}
\providecommand{\eprint}[2][]{\url{#2}}

\bibitem[{\citenamefont{Petrenko and Whitworth}(1999)}]{pe99}
\bibinfo{author}{\bibfnamefont{V.~F.} \bibnamefont{Petrenko}} \bibnamefont{and}
  \bibinfo{author}{\bibfnamefont{R.~W.} \bibnamefont{Whitworth}},
  \emph{\bibinfo{title}{Physics of Ice}} (\bibinfo{publisher}{Oxford University
  Press}, \bibinfo{address}{New York}, \bibinfo{year}{1999}).

\bibitem[{\citenamefont{Dunaeva et~al.}(2010)\citenamefont{Dunaeva, Antsyshkin,
  and Kuskov}}]{du10}
\bibinfo{author}{\bibfnamefont{A.~N.} \bibnamefont{Dunaeva}},
  \bibinfo{author}{\bibfnamefont{D.~V.} \bibnamefont{Antsyshkin}},
  \bibnamefont{and} \bibinfo{author}{\bibfnamefont{O.~L.}
  \bibnamefont{Kuskov}}, \bibinfo{journal}{Solar System Research}
  \textbf{\bibinfo{volume}{44}}, \bibinfo{pages}{202} (\bibinfo{year}{2010}).

\bibitem[{\citenamefont{Bartels-Rausch
  et~al.}(2012)\citenamefont{Bartels-Rausch, Bergeron, Cartwright, Escribano,
  Finney, Grothe, Gutierrez, Haapala, Kuhs, Pettersson et~al.}}]{ba12}
\bibinfo{author}{\bibfnamefont{T.}~\bibnamefont{Bartels-Rausch}},
  \bibinfo{author}{\bibfnamefont{V.}~\bibnamefont{Bergeron}},
  \bibinfo{author}{\bibfnamefont{J.~H.~E.} \bibnamefont{Cartwright}},
  \bibinfo{author}{\bibfnamefont{R.}~\bibnamefont{Escribano}},
  \bibinfo{author}{\bibfnamefont{J.~L.} \bibnamefont{Finney}},
  \bibinfo{author}{\bibfnamefont{H.}~\bibnamefont{Grothe}},
  \bibinfo{author}{\bibfnamefont{P.~J.} \bibnamefont{Gutierrez}},
  \bibinfo{author}{\bibfnamefont{J.}~\bibnamefont{Haapala}},
  \bibinfo{author}{\bibfnamefont{W.~F.} \bibnamefont{Kuhs}},
  \bibinfo{author}{\bibfnamefont{J.~B.~C.} \bibnamefont{Pettersson}},
  \bibnamefont{et~al.}, \bibinfo{journal}{Rev. Mod. Phys.}
  \textbf{\bibinfo{volume}{84}}, \bibinfo{pages}{885} (\bibinfo{year}{2012}).

\bibitem[{\citenamefont{Eisenberg and Kauzmann}(1969)}]{ei69}
\bibinfo{author}{\bibfnamefont{D.}~\bibnamefont{Eisenberg}} \bibnamefont{and}
  \bibinfo{author}{\bibfnamefont{W.}~\bibnamefont{Kauzmann}},
  \emph{\bibinfo{title}{The Structure and Properties of Water}}
  (\bibinfo{publisher}{Oxford University Press}, \bibinfo{address}{New York},
  \bibinfo{year}{1969}).

\bibitem[{\citenamefont{Robinson et~al.}(1996)\citenamefont{Robinson, Zhu,
  Singh, and Evans}}]{ro96}
\bibinfo{author}{\bibfnamefont{G.~W.} \bibnamefont{Robinson}},
  \bibinfo{author}{\bibfnamefont{S.~B.} \bibnamefont{Zhu}},
  \bibinfo{author}{\bibfnamefont{S.}~\bibnamefont{Singh}}, \bibnamefont{and}
  \bibinfo{author}{\bibfnamefont{M.~W.} \bibnamefont{Evans}},
  \emph{\bibinfo{title}{Water in Biology, Chemistry and Physics}}
  (\bibinfo{publisher}{World Scientific}, \bibinfo{address}{Singapore},
  \bibinfo{year}{1996}).

\bibitem[{\citenamefont{Bernal and Fowler}(1933)}]{be33}
\bibinfo{author}{\bibfnamefont{J.~D.} \bibnamefont{Bernal}} \bibnamefont{and}
  \bibinfo{author}{\bibfnamefont{R.~H.} \bibnamefont{Fowler}},
  \bibinfo{journal}{J. Chem. Phys.} \textbf{\bibinfo{volume}{1}},
  \bibinfo{pages}{515} (\bibinfo{year}{1933}).

\bibitem[{\citenamefont{Salzmann et~al.}(2011)\citenamefont{Salzmann, Radaelli,
  Slater, and Finney}}]{sa11}
\bibinfo{author}{\bibfnamefont{C.~G.} \bibnamefont{Salzmann}},
  \bibinfo{author}{\bibfnamefont{P.~G.} \bibnamefont{Radaelli}},
  \bibinfo{author}{\bibfnamefont{B.}~\bibnamefont{Slater}}, \bibnamefont{and}
  \bibinfo{author}{\bibfnamefont{J.~L.} \bibnamefont{Finney}},
  \bibinfo{journal}{Phys. Chem. Chem. Phys.} \textbf{\bibinfo{volume}{13}},
  \bibinfo{pages}{18468} (\bibinfo{year}{2011}).

\bibitem[{\citenamefont{Pauling}(1935)}]{pa35}
\bibinfo{author}{\bibfnamefont{L.}~\bibnamefont{Pauling}}, \bibinfo{journal}{J.
  Am. Chem. Soc.} \textbf{\bibinfo{volume}{57}}, \bibinfo{pages}{2680}
  (\bibinfo{year}{1935}).

\bibitem[{\citenamefont{Giauque and Stout}(1936)}]{gi36}
\bibinfo{author}{\bibfnamefont{W.~F.} \bibnamefont{Giauque}} \bibnamefont{and}
  \bibinfo{author}{\bibfnamefont{J.~W.} \bibnamefont{Stout}},
  \bibinfo{journal}{J. Amer. Chem. Soc.} \textbf{\bibinfo{volume}{58}},
  \bibinfo{pages}{1144} (\bibinfo{year}{1936}).

\bibitem[{\citenamefont{Haida et~al.}(1974)\citenamefont{Haida, Matsuo, Suga,
  and Seki}}]{ha74}
\bibinfo{author}{\bibfnamefont{O.}~\bibnamefont{Haida}},
  \bibinfo{author}{\bibfnamefont{T.}~\bibnamefont{Matsuo}},
  \bibinfo{author}{\bibfnamefont{H.}~\bibnamefont{Suga}}, \bibnamefont{and}
  \bibinfo{author}{\bibfnamefont{S.}~\bibnamefont{Seki}}, \bibinfo{journal}{J.
  Chem. Thermodyn.} \textbf{\bibinfo{volume}{6}}, \bibinfo{pages}{815}
  (\bibinfo{year}{1974}).

\bibitem[{\citenamefont{Nagle}(1966)}]{na66}
\bibinfo{author}{\bibfnamefont{J.~F.} \bibnamefont{Nagle}},
  \bibinfo{journal}{J. Math. Phys.} \textbf{\bibinfo{volume}{7}},
  \bibinfo{pages}{1484} (\bibinfo{year}{1966}).

\bibitem[{\citenamefont{Berg et~al.}(2007)\citenamefont{Berg, Muguruma, and
  Okamoto}}]{be07}
\bibinfo{author}{\bibfnamefont{B.~A.} \bibnamefont{Berg}},
  \bibinfo{author}{\bibfnamefont{C.}~\bibnamefont{Muguruma}}, \bibnamefont{and}
  \bibinfo{author}{\bibfnamefont{Y.}~\bibnamefont{Okamoto}},
  \bibinfo{journal}{Phys. Rev. B} \textbf{\bibinfo{volume}{75}},
  \bibinfo{pages}{092202} (\bibinfo{year}{2007}).

\bibitem[{\citenamefont{Berg et~al.}(2012)\citenamefont{Berg, Muguruma, and
  Okamoto}}]{be12}
\bibinfo{author}{\bibfnamefont{B.~A.} \bibnamefont{Berg}},
  \bibinfo{author}{\bibfnamefont{C.}~\bibnamefont{Muguruma}}, \bibnamefont{and}
  \bibinfo{author}{\bibfnamefont{Y.}~\bibnamefont{Okamoto}},
  \bibinfo{journal}{Mol. Sim.} \textbf{\bibinfo{volume}{38}},
  \bibinfo{pages}{856} (\bibinfo{year}{2012}).

\bibitem[{\citenamefont{Howe and Whitworth}(1987)}]{ho87}
\bibinfo{author}{\bibfnamefont{R.}~\bibnamefont{Howe}} \bibnamefont{and}
  \bibinfo{author}{\bibfnamefont{R.~W.} \bibnamefont{Whitworth}},
  \bibinfo{journal}{J. Chem. Phys.} \textbf{\bibinfo{volume}{86}},
  \bibinfo{pages}{6443} (\bibinfo{year}{1987}).

\bibitem[{\citenamefont{MacDowell et~al.}(2004)\citenamefont{MacDowell, Sanz,
  Vega, and Abascal}}]{ma04b}
\bibinfo{author}{\bibfnamefont{L.~G.} \bibnamefont{MacDowell}},
  \bibinfo{author}{\bibfnamefont{E.}~\bibnamefont{Sanz}},
  \bibinfo{author}{\bibfnamefont{C.}~\bibnamefont{Vega}}, \bibnamefont{and}
  \bibinfo{author}{\bibfnamefont{J.~L.~F.} \bibnamefont{Abascal}},
  \bibinfo{journal}{J. Chem. Phys.} \textbf{\bibinfo{volume}{121}},
  \bibinfo{pages}{10145} (\bibinfo{year}{2004}).

\bibitem[{\citenamefont{Berg and Yang}(2007)}]{be07b}
\bibinfo{author}{\bibfnamefont{B.~A.} \bibnamefont{Berg}} \bibnamefont{and}
  \bibinfo{author}{\bibfnamefont{W.}~\bibnamefont{Yang}}, \bibinfo{journal}{J.
  Chem. Phys.} \textbf{\bibinfo{volume}{127}}, \bibinfo{pages}{224502}
  (\bibinfo{year}{2007}).

\bibitem[{\citenamefont{Isakov et~al.}(2004)\citenamefont{Isakov, Raman,
  Moessner, and Sondhi}}]{is04}
\bibinfo{author}{\bibfnamefont{S.~V.} \bibnamefont{Isakov}},
  \bibinfo{author}{\bibfnamefont{K.~S.} \bibnamefont{Raman}},
  \bibinfo{author}{\bibfnamefont{R.}~\bibnamefont{Moessner}}, \bibnamefont{and}
  \bibinfo{author}{\bibfnamefont{S.~L.} \bibnamefont{Sondhi}},
  \bibinfo{journal}{Phys. Rev. B} \textbf{\bibinfo{volume}{70}},
  \bibinfo{pages}{104418} (\bibinfo{year}{2004}).

\bibitem[{\citenamefont{Pomaranski et~al.}(2013)\citenamefont{Pomaranski,
  Yaraskavitch, Meng, Ross, Noad, Dabkowska, Gaulin, and Kycia}}]{po13}
\bibinfo{author}{\bibfnamefont{D.}~\bibnamefont{Pomaranski}},
  \bibinfo{author}{\bibfnamefont{L.~R.} \bibnamefont{Yaraskavitch}},
  \bibinfo{author}{\bibfnamefont{S.}~\bibnamefont{Meng}},
  \bibinfo{author}{\bibfnamefont{K.~A.} \bibnamefont{Ross}},
  \bibinfo{author}{\bibfnamefont{H.~M.~L.} \bibnamefont{Noad}},
  \bibinfo{author}{\bibfnamefont{H.~A.} \bibnamefont{Dabkowska}},
  \bibinfo{author}{\bibfnamefont{B.~D.} \bibnamefont{Gaulin}},
  \bibnamefont{and} \bibinfo{author}{\bibfnamefont{J.~B.} \bibnamefont{Kycia}},
  \bibinfo{journal}{Nature Phys.} \textbf{\bibinfo{volume}{9}},
  \bibinfo{pages}{353} (\bibinfo{year}{2013}).

\bibitem[{\citenamefont{Ziman}(1979)}]{zi79}
\bibinfo{author}{\bibfnamefont{J.~M.} \bibnamefont{Ziman}},
  \emph{\bibinfo{title}{Models of disorder}} (\bibinfo{publisher}{Cambridge
  University}, \bibinfo{address}{Cambridge}, \bibinfo{year}{1979}).

\bibitem[{\citenamefont{Bell and Lavis}(1989)}]{be89}
\bibinfo{author}{\bibfnamefont{G.~M.} \bibnamefont{Bell}} \bibnamefont{and}
  \bibinfo{author}{\bibfnamefont{D.~A.} \bibnamefont{Lavis}},
  \emph{\bibinfo{title}{Statistical Mechanics of Lattice Models. Volume 1:
  Closed Form and Exact Theories of Cooperative Phenomena}}
  (\bibinfo{publisher}{Ellis Horwood Ltd.}, \bibinfo{address}{New York},
  \bibinfo{year}{1989}).

\bibitem[{\citenamefont{Lieb}(1967{\natexlab{a}})}]{li67}
\bibinfo{author}{\bibfnamefont{E.~H.} \bibnamefont{Lieb}},
  \bibinfo{journal}{Phys. Rev. Lett.} \textbf{\bibinfo{volume}{18}},
  \bibinfo{pages}{692} (\bibinfo{year}{1967}{\natexlab{a}}).

\bibitem[{\citenamefont{Lieb}(1967{\natexlab{b}})}]{li67b}
\bibinfo{author}{\bibfnamefont{E.~H.} \bibnamefont{Lieb}},
  \bibinfo{journal}{Phys. Rev.} \textbf{\bibinfo{volume}{162}},
  \bibinfo{pages}{162} (\bibinfo{year}{1967}{\natexlab{b}}).

\bibitem[{\citenamefont{Herrero and Ram\'irez}(2013{\natexlab{a}})}]{he13}
\bibinfo{author}{\bibfnamefont{C.~P.} \bibnamefont{Herrero}} \bibnamefont{and}
  \bibinfo{author}{\bibfnamefont{R.}~\bibnamefont{Ram\'irez}},
  \bibinfo{journal}{Chem. Phys. Lett.} \textbf{\bibinfo{volume}{568-569}},
  \bibinfo{pages}{70} (\bibinfo{year}{2013}{\natexlab{a}}).

\bibitem[{\citenamefont{Knight et~al.}(2006)\citenamefont{Knight, Singer, Kuo,
  Hirsch, Ojamae, and Klein}}]{kn06}
\bibinfo{author}{\bibfnamefont{C.}~\bibnamefont{Knight}},
  \bibinfo{author}{\bibfnamefont{S.~J.} \bibnamefont{Singer}},
  \bibinfo{author}{\bibfnamefont{J.~L.} \bibnamefont{Kuo}},
  \bibinfo{author}{\bibfnamefont{T.~K.} \bibnamefont{Hirsch}},
  \bibinfo{author}{\bibfnamefont{L.}~\bibnamefont{Ojamae}}, \bibnamefont{and}
  \bibinfo{author}{\bibfnamefont{M.~L.} \bibnamefont{Klein}},
  \bibinfo{journal}{Phys. Rev. E} \textbf{\bibinfo{volume}{75}},
  \bibinfo{pages}{056113} (\bibinfo{year}{2006}).

\bibitem[{\citenamefont{Singer and Knight}(2012)}]{si12}
\bibinfo{author}{\bibfnamefont{S.~J.} \bibnamefont{Singer}} \bibnamefont{and}
  \bibinfo{author}{\bibfnamefont{C.}~\bibnamefont{Knight}},
  \bibinfo{journal}{Adv. Chem. Phys.} \textbf{\bibinfo{volume}{147}},
  \bibinfo{pages}{1} (\bibinfo{year}{2012}).

\bibitem[{\citenamefont{Binder and Heermann}(2010)}]{bi10}
\bibinfo{author}{\bibfnamefont{K.}~\bibnamefont{Binder}} \bibnamefont{and}
  \bibinfo{author}{\bibfnamefont{D.~W.} \bibnamefont{Heermann}},
  \emph{\bibinfo{title}{Monte Carlo Simulation in Statistical Physics}}
  (\bibinfo{publisher}{Springer}, \bibinfo{address}{Berlin},
  \bibinfo{year}{2010}), \bibinfo{edition}{5th} ed.

\bibitem[{\citenamefont{Chandler}(1987)}]{ch87}
\bibinfo{author}{\bibfnamefont{D.}~\bibnamefont{Chandler}},
  \emph{\bibinfo{title}{Introduction to modern statistical mechanics}}
  (\bibinfo{publisher}{Oxford University Press}, \bibinfo{address}{Oxford},
  \bibinfo{year}{1987}).

\bibitem[{\citenamefont{Herrero and Ram\'{\i}rez}(1992)}]{he92}
\bibinfo{author}{\bibfnamefont{C.~P.} \bibnamefont{Herrero}} \bibnamefont{and}
  \bibinfo{author}{\bibfnamefont{R.}~\bibnamefont{Ram\'{\i}rez}},
  \bibinfo{journal}{Chem. Phys. Lett.} \textbf{\bibinfo{volume}{194}},
  \bibinfo{pages}{79} (\bibinfo{year}{1992}).

\bibitem[{\citenamefont{Pearson}(1972)}]{pe72}
\bibinfo{author}{\bibfnamefont{W.~B.} \bibnamefont{Pearson}},
  \emph{\bibinfo{title}{The crystal chemistry and physics of metals and
  alloys}} (\bibinfo{publisher}{Wiley}, \bibinfo{address}{New York},
  \bibinfo{year}{1972}).

\bibitem[{\citenamefont{Liebau}(1985)}]{li85}
\bibinfo{author}{\bibfnamefont{F.}~\bibnamefont{Liebau}},
  \emph{\bibinfo{title}{Structural chemistry of silicates: structure, bonding,
  and classification}} (\bibinfo{publisher}{Springer},
  \bibinfo{address}{Berlin}, \bibinfo{year}{1985}).

\bibitem[{\citenamefont{Wells}(1986)}]{we86}
\bibinfo{author}{\bibfnamefont{A.~F.} \bibnamefont{Wells}},
  \emph{\bibinfo{title}{Structural inorganic chemistry}}
  (\bibinfo{publisher}{Clarendon Press}, \bibinfo{address}{Oxford},
  \bibinfo{year}{1986}), \bibinfo{edition}{5th} ed.

\bibitem[{\citenamefont{Herrero and Ram\'irez}(2013{\natexlab{b}})}]{he13b}
\bibinfo{author}{\bibfnamefont{C.~P.} \bibnamefont{Herrero}} \bibnamefont{and}
  \bibinfo{author}{\bibfnamefont{R.}~\bibnamefont{Ram\'irez}},
  \bibinfo{journal}{Phys. Chem. Chem. Phys.} \textbf{\bibinfo{volume}{15}},
  \bibinfo{pages}{16676} (\bibinfo{year}{2013}{\natexlab{b}}).

\bibitem[{\citenamefont{Domb}(1970)}]{do70}
\bibinfo{author}{\bibfnamefont{C.}~\bibnamefont{Domb}}, \bibinfo{journal}{J.
  Phys. C: Solid State Phys.} \textbf{\bibinfo{volume}{3}},
  \bibinfo{pages}{256} (\bibinfo{year}{1970}).

\bibitem[{\citenamefont{Herrero}(1995)}]{he95b}
\bibinfo{author}{\bibfnamefont{C.~P.} \bibnamefont{Herrero}},
  \bibinfo{journal}{J. Phys.: Condens. Matter} \textbf{\bibinfo{volume}{7}},
  \bibinfo{pages}{8897} (\bibinfo{year}{1995}).

\bibitem[{\citenamefont{Herrero}(2014)}]{he14}
\bibinfo{author}{\bibfnamefont{C.~P.} \bibnamefont{Herrero}},
  \bibinfo{journal}{Chem. Phys.} \textbf{\bibinfo{volume}{439}},
  \bibinfo{pages}{49} (\bibinfo{year}{2014}).

\bibitem[{\citenamefont{McKenzie}(1976)}]{mc76}
\bibinfo{author}{\bibfnamefont{D.~S.} \bibnamefont{McKenzie}},
  \bibinfo{journal}{Phys. Rep.} \textbf{\bibinfo{volume}{27}},
  \bibinfo{pages}{35} (\bibinfo{year}{1976}).

\bibitem[{\citenamefont{Rapaport}(1985)}]{ra85}
\bibinfo{author}{\bibfnamefont{D.~C.} \bibnamefont{Rapaport}},
  \bibinfo{journal}{J. Phys. A: Math. Gen.} \textbf{\bibinfo{volume}{18}},
  \bibinfo{pages}{113} (\bibinfo{year}{1985}).

\bibitem[{\citenamefont{Privman et~al.}(1991)\citenamefont{Privman, Hohenberg,
  and Aharoni}}]{pr91}
\bibinfo{author}{\bibfnamefont{V.}~\bibnamefont{Privman}},
  \bibinfo{author}{\bibfnamefont{P.~C.} \bibnamefont{Hohenberg}},
  \bibnamefont{and} \bibinfo{author}{\bibfnamefont{A.}~\bibnamefont{Aharoni}},
  in \emph{\bibinfo{booktitle}{Phase Transitions and Critical Phenomena}},
  edited by \bibinfo{editor}{\bibfnamefont{C.}~\bibnamefont{Domb}}
  \bibnamefont{and} \bibinfo{editor}{\bibfnamefont{J.~L.}
  \bibnamefont{Lebowitz}} (\bibinfo{publisher}{Academic Press},
  \bibinfo{address}{London}, \bibinfo{year}{1991}), vol.~\bibinfo{volume}{14},
  pp. \bibinfo{pages}{1--134}.

\bibitem[{\citenamefont{Caracciolo et~al.}(1998)\citenamefont{Caracciolo,
  Causo, and Pelissetto}}]{ca98}
\bibinfo{author}{\bibfnamefont{S.}~\bibnamefont{Caracciolo}},
  \bibinfo{author}{\bibfnamefont{M.~S.} \bibnamefont{Causo}}, \bibnamefont{and}
  \bibinfo{author}{\bibfnamefont{A.}~\bibnamefont{Pelissetto}},
  \bibinfo{journal}{Phys. Rev. E} \textbf{\bibinfo{volume}{57}},
  \bibinfo{pages}{R1215} (\bibinfo{year}{1998}).

\bibitem[{\citenamefont{Chen and Lin}(2002)}]{ch02}
\bibinfo{author}{\bibfnamefont{M.}~\bibnamefont{Chen}} \bibnamefont{and}
  \bibinfo{author}{\bibfnamefont{K.~Y.} \bibnamefont{Lin}},
  \bibinfo{journal}{J. Phys. A: Math. Gen.} \textbf{\bibinfo{volume}{35}},
  \bibinfo{pages}{1501} (\bibinfo{year}{2002}).

\bibitem[{\citenamefont{Herrero}(2002)}]{he02b}
\bibinfo{author}{\bibfnamefont{C.~P.} \bibnamefont{Herrero}},
  \bibinfo{journal}{Phys. Rev. E} \textbf{\bibinfo{volume}{66}},
  \bibinfo{pages}{046126} (\bibinfo{year}{2002}).

\bibitem[{\citenamefont{Herrero and Saboy\'a}(2003)}]{he03c}
\bibinfo{author}{\bibfnamefont{C.~P.} \bibnamefont{Herrero}} \bibnamefont{and}
  \bibinfo{author}{\bibfnamefont{M.}~\bibnamefont{Saboy\'a}},
  \bibinfo{journal}{Phys. Rev. E} \textbf{\bibinfo{volume}{68}},
  \bibinfo{pages}{026106} (\bibinfo{year}{2003}).

\bibitem[{\citenamefont{Pelissetto and Vicari}(2002)}]{pe02}
\bibinfo{author}{\bibfnamefont{A.}~\bibnamefont{Pelissetto}} \bibnamefont{and}
  \bibinfo{author}{\bibfnamefont{E.}~\bibnamefont{Vicari}},
  \bibinfo{journal}{Phys. Rep.} \textbf{\bibinfo{volume}{368}},
  \bibinfo{pages}{549} (\bibinfo{year}{2002}).

\bibitem[{\citenamefont{Hollins}(1964)}]{ho64}
\bibinfo{author}{\bibfnamefont{G.~T.} \bibnamefont{Hollins}},
  \bibinfo{journal}{Proc. Phys. Soc.} \textbf{\bibinfo{volume}{84}},
  \bibinfo{pages}{1001} (\bibinfo{year}{1964}).

\bibitem[{\citenamefont{Sanz et~al.}(2004)\citenamefont{Sanz, Vega, Abascal,
  and MacDowell}}]{sa04}
\bibinfo{author}{\bibfnamefont{E.}~\bibnamefont{Sanz}},
  \bibinfo{author}{\bibfnamefont{C.}~\bibnamefont{Vega}},
  \bibinfo{author}{\bibfnamefont{J.~L.~F.} \bibnamefont{Abascal}},
  \bibnamefont{and} \bibinfo{author}{\bibfnamefont{L.~G.}
  \bibnamefont{MacDowell}}, \bibinfo{journal}{Phys. Rev. Lett.}
  \textbf{\bibinfo{volume}{92}}, \bibinfo{pages}{255701}
  (\bibinfo{year}{2004}).

\bibitem[{\citenamefont{Ram\'{\i}rez et~al.}(2012)\citenamefont{Ram\'{\i}rez,
  Neuerburg, and Herrero}}]{ra12b}
\bibinfo{author}{\bibfnamefont{R.}~\bibnamefont{Ram\'{\i}rez}},
  \bibinfo{author}{\bibfnamefont{N.}~\bibnamefont{Neuerburg}},
  \bibnamefont{and} \bibinfo{author}{\bibfnamefont{C.~P.}
  \bibnamefont{Herrero}}, \bibinfo{journal}{J. Chem. Phys.}
  \textbf{\bibinfo{volume}{137}}, \bibinfo{pages}{134503}
  (\bibinfo{year}{2012}).

\bibitem[{\citenamefont{Ram\'irez et~al.}(2013)\citenamefont{Ram\'irez,
  Neuerburg, and Herrero}}]{ra13}
\bibinfo{author}{\bibfnamefont{R.}~\bibnamefont{Ram\'irez}},
  \bibinfo{author}{\bibfnamefont{N.}~\bibnamefont{Neuerburg}},
  \bibnamefont{and} \bibinfo{author}{\bibfnamefont{C.~P.}
  \bibnamefont{Herrero}}, \bibinfo{journal}{J. Chem. Phys.}
  \textbf{\bibinfo{volume}{139}}, \bibinfo{pages}{084503}
  (\bibinfo{year}{2013}).

\bibitem[{\citenamefont{Peterson and Levy}(1957)}]{pe57}
\bibinfo{author}{\bibfnamefont{S.~W.} \bibnamefont{Peterson}} \bibnamefont{and}
  \bibinfo{author}{\bibfnamefont{H.~A.} \bibnamefont{Levy}},
  \bibinfo{journal}{Acta Cryst.} \textbf{\bibinfo{volume}{10}},
  \bibinfo{pages}{70} (\bibinfo{year}{1957}).

\bibitem[{\citenamefont{K\"onig}(1944)}]{ko44}
\bibinfo{author}{\bibfnamefont{H.}~\bibnamefont{K\"onig}}, \bibinfo{journal}{Z.
  Kristallogr.} \textbf{\bibinfo{volume}{105}}, \bibinfo{pages}{279}
  (\bibinfo{year}{1944}).

\bibitem[{\citenamefont{Kamb et~al.}(1971)\citenamefont{Kamb, Hamilton,
  {LaPlaca}, and Prakash}}]{ka71}
\bibinfo{author}{\bibfnamefont{B.}~\bibnamefont{Kamb}},
  \bibinfo{author}{\bibfnamefont{W.~C.} \bibnamefont{Hamilton}},
  \bibinfo{author}{\bibfnamefont{S.~J.} \bibnamefont{{LaPlaca}}},
  \bibnamefont{and} \bibinfo{author}{\bibfnamefont{A.}~\bibnamefont{Prakash}},
  \bibinfo{journal}{J. Chem. Phys.} \textbf{\bibinfo{volume}{55}},
  \bibinfo{pages}{1934} (\bibinfo{year}{1971}).

\bibitem[{\citenamefont{Lobban et~al.}(2000)\citenamefont{Lobban, Finney, and
  Kuhs}}]{lo00}
\bibinfo{author}{\bibfnamefont{C.}~\bibnamefont{Lobban}},
  \bibinfo{author}{\bibfnamefont{J.~L.} \bibnamefont{Finney}},
  \bibnamefont{and} \bibinfo{author}{\bibfnamefont{W.~F.} \bibnamefont{Kuhs}},
  \bibinfo{journal}{J. Chem. Phys.} \textbf{\bibinfo{volume}{112}},
  \bibinfo{pages}{7169} (\bibinfo{year}{2000}).

\bibitem[{\citenamefont{Engelhardt and Kamb}(1981)}]{en81}
\bibinfo{author}{\bibfnamefont{H.}~\bibnamefont{Engelhardt}} \bibnamefont{and}
  \bibinfo{author}{\bibfnamefont{B.}~\bibnamefont{Kamb}}, \bibinfo{journal}{J.
  Chem. Phys.} \textbf{\bibinfo{volume}{75}}, \bibinfo{pages}{5887}
  (\bibinfo{year}{1981}).

\bibitem[{\citenamefont{Kuhs et~al.}(1984)\citenamefont{Kuhs, Finney, Vettier,
  and Bliss}}]{ku84}
\bibinfo{author}{\bibfnamefont{W.~F.} \bibnamefont{Kuhs}},
  \bibinfo{author}{\bibfnamefont{J.~L.} \bibnamefont{Finney}},
  \bibinfo{author}{\bibfnamefont{C.}~\bibnamefont{Vettier}}, \bibnamefont{and}
  \bibinfo{author}{\bibfnamefont{D.~V.} \bibnamefont{Bliss}},
  \bibinfo{journal}{J. Chem. Phys.} \textbf{\bibinfo{volume}{81}},
  \bibinfo{pages}{3612} (\bibinfo{year}{1984}).

\bibitem[{\citenamefont{Lobban et~al.}(1998)\citenamefont{Lobban, Finney, and
  Kuhs}}]{lo98}
\bibinfo{author}{\bibfnamefont{C.}~\bibnamefont{Lobban}},
  \bibinfo{author}{\bibfnamefont{J.~L.} \bibnamefont{Finney}},
  \bibnamefont{and} \bibinfo{author}{\bibfnamefont{W.~F.} \bibnamefont{Kuhs}},
  \bibinfo{journal}{Nature} \textbf{\bibinfo{volume}{391}},
  \bibinfo{pages}{268} (\bibinfo{year}{1998}).

\end{thebibliography}
\end{document}